%
\input phyzzx
\hfuzz 20pt
\font\mybb=msbm10 at 12pt

\def\Bbb#1{\hbox{\mybb#1}}

\def\bC {\Bbb{C}}

\def\bE{\Bbb {E}}

\def\bZ {\Bbb{Z}}
\def\bfomega{\omega\kern-7.0pt \omega}


\REF\gppkt{G. Papadopoulos \& P.K. Townsend,  
{\sl Intersecting M-branes}, Phys. Lett.
 {\bf B380} (1996) 273.}
\REF\strom{A. Strominger, {\sl Open p-branes}, Phys. Lett. 
{\bf B383} 44; hep-th/9512059.}
\REF\pt{P.K. Townsend, {\sl Brane Surgery}, Nucl. 
Phys. Proc. Suppl {\bf 58} (1997) 163.}
\REF\malda{C.G. Callan, Jr \& J.M. Maldacena, 
{\sl Brane dynamics from the
Born-Infeld action}, hep-th/9708147. }
\REF\gary{G.W. Gibbons, {\sl Born-Infeld particles 
and Dirichlet p-branes}, hep-th/9709027.}
\REF\garyb{G.W. Gibbons, {\sl Wormholes on the 
Worldvolume: Born-Infeld Particles
and Dirichlet p-Branes}, hep-th/9801106.}
\REF\green{M.B. Green \& J.H. Schwarz, {\sl Covariant 
Description of Superstrings}, Phys.
Lett. {\bf 136B} (1984) 367.}
\REF\bst{ E. Bergshoeff, E. Sezgin \& P.K. 
Townsend, {\sl Supermembranes and
Eleven-Dimensional Supergravity}, Phys. Lett. 
{\bf 189B} (1987) 75.}
\REF\refda{M. Cederwall, A. von Gussich, B.E.W.
 Nilsson, P. Sundell \& A. Westerberg, 
{\sl The Dirichlet Super P-Branes in Ten-Dimensional
 Type IIA and IIB Supergravity}, Nucl.
Phys. {\bf B490} (1997) 179.}
\REF\refdb{M. Aganagic, C. Popescu \& J.H. Schwarz, 
{\sl D-Brane Actions with Local
Kappa Symmetry}, Phys. Lett. {\bf B393} (1997) 311.}
\REF\refdc{ E. Bergshoeff \& P.K. Townsend, 
{\sl Super D-Branes}, Nucl. Phys. {\bf B490}
(1997) 145.}
\REF\refma{ P.S. Howe \& E. Sezgin, 
{\sl D=11, p=5}, Phys. Lett. {\bf B394} (1997) 62.}
\REF\refmb{ I. Bandos, K. Lechner,
A. Nurmagambetov, P.Pasti, D. Sorokin \& M. Tonin, {\sl
Covariant Action for the Super-Five-Brane of M-theory}, 
hep-th/9703127.}
\REF\refmc{M. Aganagic, J. Park, C. Popescu 
\& J.H. Schwarz, {\sl Worldvolume Action of the
M-Theory Five-Brane}, hep-th/9701166.}
\REF\cali {R. Harvey and H. Blaine Lawson, Jr,
{\sl Calibrated Geometries} Acta Math. {\bf
148} (1982) 47.}
\REF\calib{F.R. Harvey, {\sl Spinors and Calibrations}, 
Academic Press (1990), New York.}
\REF\beckera{K. Becker, M. Becker \& A. Strominger, 
{\sl Fivebranes, Membranes and
Nonperturbative String Theory}, Nucl. Phys. 
{\bf B456} (1995) 130.}
\REF\beckerb{K. Becker, M. Becker,
 D.R. Morrison, H. Ooguri, Y. Oz \& Z. Yin, {\sl
Supersymmetric Cycles in Exceptional Holonomy
 Manifolds and Calabi-Yau 4-folds},
hep-th/9608116.}
\REF\douglas{ M.R. Douglas, {\sl Branes within Branes}, hep-th/9512077.}
\REF\gpgpa{J.M. Izquierdo, N.D. Lambert, 
G. Papadopoulos \& P.K. Townsend, {\sl Dyonic
Branes}, Nucl. Phys. {\bf B460} (1996) 560.}
\REF\gpgpb{G. Papadopoulos \& P.K. Townsend, 
{\sl Kaluza-Klein on the brane}, Phys. Lett.
{\bf B393} (1997) 59.}
\REF\kallosh {E. Bergshoeff, R. Kallosh, G. 
Papadopoulos \& T. Ortin, {\sl kappa-symmetry
supersymmetry and intersecting branes}, Nucl. Phys. {\bf B502} (1997) 149.}
\REF\douglasb{M. Berkooz, M.R. Douglas and 
R.G. Leigh, {\sl Branes Intersecting at Angles},
Nucl. Phys. {\bf B480} (1996) 265.}
\REF\garypaul{J. P. Gauntlett, G.W. Gibbons, 
G. Papadopoulos \& P.K. Townsend, {\sl
Hyper-K\"ahler Manifolds and Multiply 
Intersecting Branes}, Nucl. Phys. {\bf B500} (1997)
133.} 
\REF\wang{McKenzie Y.Wang, {\sl Parallel Spinors and Parallel Forms}, 
Ann. Global Anal.
Geom. {\bf 7} (1989), 59.}
\REF\donald{S.K. Donaldson, {\sl Anti Self-dual 
Yang-Mills Connections over Complex
Algebraic Surfaces and Stable Bundles  }, Proc. 
London Math. Soc. {\bf 50} (1985) 1.}
\REF\kent{E. Corrigan, P. Goddard \& A. Kent, 
{\sl Some Comments on the ADHM Construction
in 4k Dimensions}, Commun. Math. Phys. {\bf 100} (1985) 1.}
\REF\fn{ S. Fubini \& H. Nicolai, {\sl The 
Octonionic Instanton}, Phys. Lett. {\bf 155B}
(1985) 369.}
\REF\ivanova{T.A. Ivanova, {\sl Octonions, 
Self-Duality and Strings}, Phys. Lett. {\bf
B315} (1993) 277.}
\REF\gn{M. Gunaydin \& H. Nicolai, {\sl Seven-Dimensional
 Octonionic Yang-Mills Instanton
and its Extension to an Heterotic String Soliton}, 
Phys. Lett. {\bf B351} (1995) 169.}
\REF\gptf{ G. Papadopoulos \& A. Teschendorff, 
{\sl Instantons at Angles}, hep-th/9708116.}
\REF\tseytlinb{A.A. Tseytlin, {\sl Harmonic 
Superposition of M-Branes}, Nucl. Phys.{\bf B475}
(1996) 149.}
\REF\kastor{J.P. Gauntlett, D.A. Kastor \& J. 
Traschen, {\sl Overlapping Branes in M-theory}
{\sl Nucl. Phys. } {\bf B478} (1996) 544.}
 \REF\witten{E.Witten {\sl Solutions of four-dimensional 
field theories via M-theory}, Nucl. Phys.
{\bf B500} (1997) 3.}
\REF\schwarzb{J.H. Schwarz, {\sl Lectures on 
Superstrings and M-theory Dualities}, 
Nucl. Phys. Proc. Suppl.{\bf 55B} (1997) 1.}
\REF\sena{A. Sen, {\sl String Network}, hep-th/9711130.}
\REF\krogh{ M. Krogh \& S. Lee, {\sl String 
Network from M-theory}, hep-th/9712050.}
\REF\rey{Soo-Jong Rey\& Jung-Tay Yee, {\sl BPS Dynamics of Triple
(p,q) String Junction}, hep-th/9711202.}
\REF\howeb{P.S. Howe, N.D. Lambert and P. West, 
{\sl The self-dual string soliton},
hep-th/9709014.}
\REF\gpapas{G. Papadopoulos, {\sl T-duality and 
the worldvolume solitons of five-branes
and KK-monopoles}, hep-th/9712162.}
\REF\bergpap{E. Bergshoeff, J.P. van der Schaar 
\& G. Papadopoulos, {\sl Domain Walls
on the Brane}, hep-th/9801158.}
\REF\gomis{J. Gauntlett, J. Gomis \& P.K. 
Townsend, {\sl BPS Bounds for Worldvolume Branes}, hep-th/9711205.}
\REF\gpapasb{G. Papadopoulos \& J. Gutowski, 
{\sl The moduli spaces of worldvolume brane
solitons}, hep-th/9802186.}
\REF\tseytlin{A.A. Tseytlin, {\sl A Non-Abelian 
Generalization of Born-Infeld Action
in String Theory}, Nucl. Phys. {\bf B501} (1997) 41.}

\Pubnum{ \vbox{ \hbox{R/98/16}\hbox{} } }
\pubtype{}
\date{March, 1998}
\titlepage
\title{Calibrations and Intersecting Branes}
\author{G.W. Gibbons and G. Papadopoulos}
\address{DAMTP,\break Silver Street, \break  
University of Cambridge,\break Cambridge CB3
9EW}
\abstract {We investigate the solutions of 
Nambu-Goto-type  actions associated
with calibrations. We determine the 
supersymmetry preserved by these solutions
using the  contact set of the calibration and
 examine their bulk interpretation as
intersecting branes.  We show that the 
supersymmetry preserved by such solutions
is closely related to the spinor singlets
 of the subgroup $G$ of ${\rm Spin} (9,1)$ or ${\rm Spin} (10,1)$ that
rotates the tangent spaces of the brane. 
We find that
the supersymmetry projections of the worldvolume
 solutions are precisely those of the
associated bulk configurations. We also 
investigate the supersymmetric solutions of a
Born-Infeld action. We show that in some 
cases this problem also  reduces to 
counting spinor singlets of a subgroup 
of ${\rm Spin}(9,1)$ acting on the associated spinor
representations. We also find new 
worldvolume solutions   which preserve
$1/8$ of the supersymmetry of the
 bulk and give their bulk interpretation. }

\endpage
\pagenumber=2



\def\C{\mkern1mu\raise2.2pt\hbox{$\scriptscriptstyle|$}\mkern-7mu{\rm C}}



\chapter{ Introduction}

Many of the insights in the relations
 amongst the superstring theories and M-theory
have been found by investigated the 
soliton-like solutions of the ten- and
eleven-dimensional supergravity theories.  Most 
attention so far has been concentrated on 
supersymmetric solutions, {\sl i.e.} 
those that they preserve
a proportion of the supersymmetry of the
 underlying theory. Some of the supersymmetric
solutions also saturate a Bogomol'nyi bound and so they are BPS. It is
remarkable that a large class of 
supersymmetric solutions of the supergravity theories 
can be constructed by superposing
elementary soliton solutions which preserve 1/2 of the
 spacetime supersymmetry [\gppkt].  These elementary
solutions are the various brane solutions of
supergravity theories, the pp-wave and KK-monopole.
After such a superposition, the resulting solutions have the 
interpretation of intersecting branes or 
branes ending on other branes [\strom,\gppkt,\pt],
and whenever appropriate, in the
 background of a pp-wave or a KK-monopole.
The  small fluctuations of the (elementary)
 p-branes of superstrings and M-theory
are described by  (p+1)-dimensional  
worldvolume actions of Dirac-Born-Infeld   type.  
Recently, the soliton-like solutions of these
 worldvolume actions have been investigated
[\malda, \gary, \garyb]. It has been found that 
 some of these, so called worldvolume,
solutions viewed from the bulk (supergravity) 
perspective also have the interpretation of
intersecting branes or branes ending on other branes. 

One way to explain this correspondence between the
 supergravity solutions
and those of the worldvolume actions is to consider
 the supergravity coupled to the worldvolume action of a D-p-brane.
 The latter is a (p+1)-dimensional Dirac-Born-Infeld (DBI)   action.  Such an
action can be written schematically as
$$
S={1\over g^2_s} S_{NS\otimes NS}+ S_{R\otimes R}+ {1\over g_s} S_{DBI}
\eqn\inone
$$
where $g_s$ is the string coupling constant.  
The first two terms are
 the NS$\otimes$NS and R$\otimes$R parts of the
supergravity action, respectively, while 
the last term is the  Born-Infeld
 action for a D-brane propagating in a
supergravity background. In the limit that
 the string coupling 
constant becomes very large,  $S_{NS\otimes NS}$ 
diverges faster. In order to keep 
the action small, we set $S_{NS\otimes NS}=0$ which can be 
achieved by taking flat spacetime
background and setting the  
field strength of the fundamental string equal to zero.  
 This is the
weak coupling limit which can be described by 
string perturbation theory. 
Note that the $S_{R\otimes R}$ remains in this limit; 
it can be eliminated though by choosing
the D-brane field strengths to be zero.  Next suppose
 that the string coupling becomes large.
In this limit using a similar reasoning, one can
 set $S_{DBI}=0$ leading to the usual
supergravity (bulk) description of the various branes. 
 Although these limits are not consistent truncations,
it is expected that the various BPS configurations 
survive in the various limits of the string
coupling constant which explains the presence of 
intersecting brane configurations as
solutions of the Dirac-Born-Infeld action. Moreover
 the worldvolume solitons of
other branes are related to those of D-branes 
by T-and S-duality transformations. Therefore,
the above correspondence between bulk 
configurations and worldvolume ones is valid for all
branes.

The worldvolume actions of branes are 
described by matter, vector and tensor
multiplets.
For convenience we shall refer to the vectors
 and tensors fields of the vector and tensor
multiplets as Born-Infeld  fields. The 
worldvolume actions of branes described by matter
multiplets are of Nambu-Goto type. Such 
actions include those of the IIA string
and the M-2-brane [\green,\bst]. The 
worldvolume actions of  branes described by
vector and tensor multiplets for vanishing
 Born-Infeld fields can be consistently truncated
to  Nambu-Goto ones. Such actions include 
those of D-branes [\refda, \refdb, \refdc],
M-5-brane [\refma, \refmb, \refmc] and 
NS-5-brane of IIB theory. Therefore solutions of the
worldvolume theories of all branes which  do not involve
Born-Infeld fields are solutions of  
Nambu-Goto actions.  The solutions of Nambu-Goto action
are minimal surfaces.  A large class of 
minimal surfaces is that constructed using
calibrations [\cali, \calib]. Such solutions
saturate a bound, so they are BPS, and they preserve a
proportion of the spacetime supersymmetry.  The supersymmetry of
 calibrations in Calabi-Yau manifolds and 
in manifolds of exceptional holonomy  has been
investigated in [\beckera, \beckerb]. The bulk 
interpretation of the calibrated worldvolume
solutions is that  of intersecting branes.  
Another consistent truncation of the worldvolume
actions of branes is to set all the matter fields
 of vector and tensor multiplets to be
constant. The resulting action for vector 
multiplets is that of non-linear electro-dynamics,
{\sl i.e.} the Born-Infeld action.  The bulk 
interpretation of the solutions of worldvolume
theories,  that involve only Born-Infeld 
fields is that of branes within branes
[\douglas] (see [\gpgpa,\gpgpb] for the 
supergravity solutions).  The associated  bound
states can be at or below threshold. Finally, 
there are worldvolume solutions that involve
both Born-Infeld and matter fields. Such 
solutions have the bulk interpretation of branes
ending at branes. The presence of Born-Infeld fields 
is then required by the Gauss law. We
remark that  all Dirac-Born-Infeld actions relevant 
to branes, apart from that of the
M-5-brane, can be constructed by dimensionaly 
reducing the ten-dimensional Born-Infeld action
to an appropriate dimension.  Therefore all 
solutions of the Dirac-Born-Infeld actions can be
thought as solutions of the ten-dimensional Born-Infeld action. In fact
one has the striking geometric property that see e.g. [\gary]
{\it any calibrated $p$-dimensional submanifold of ${\Bbb E}^n$ may be
 regarded as
a solution of the Born-Infeld action for a $U(1)$
gauge field in ${\Bbb E}^{n+p}$ }.

In this paper, we shall investigate the solutions 
of Nambu-Goto actions associated with
calibrations. We shall first show that all these 
solutions are supersymmetric. Our proof
is
in three steps 

\item{(i)} we shall use the  Killing spinor 
equations [\kallosh] which are  derived
from the  kappa-symmetry transformations  of
 brane worldvolume actions, 
\item{(ii)} we shall
exploit the fact that the collection of the 
tangent spaces of a calibration span a subspace of
the contact set, and 
\item{(iii)} we shall apply the notion
of branes at angles [\douglasb, \garypaul]. 

The main point of the proof is that the contact set of a
calibration is a subspace of a homogeneous space, $G/H$, 
and that $G$ leaves one-dimensional
invariant subspaces (singlets) when acting 
on  the spinor representations of
${\rm Spin}(9,1)$ (or ${\rm Spin}(10,1)$) $\supset G$. 
 The number of such 
singlets determines the proportion of the
supersymmetry preserved. The groups $G$ 
that arise in calibrations are some of those that
appear in the context of special holonomies,
 {\sl i.e.} $G$ is $SU(n)$, $G_2$ and
${\rm Spin}(7)$.{\sl Our approach provides  
an alternative way to define calibrations in terms
of spinors as opposed to their usual 
definition in terms of forms.}   Then we shall
systematically explore the bulk interpretation of 
all the calibrated solutions of Nambu-Goto
actions. We shall find that in most cases there is a
supergravity intersecting brane configuration
 associated with every worldvolume solution.
Moreover we shall show that all the the 
supersymmetry projections associated with a bulk
configuration are precisely those that arise 
in the corresponding worldvolume solution. Next
we shall investigate a certain class of 
singular solutions which may arise as limits of
regular ones and shall examine the 
supersymmetry preserved by such solutions. Then we shall
investigate the supersymmetry preserved by 
worldvolume solutions which involve only a
non-vanishing  Born-Infeld field. The term in 
the Killing spinor equation that involves the
Born-Infeld field can be interpreted as spinor 
rotation [\kallosh]. Using this, we shall find
that the supersymmetry condition on the 
Born-Infeld field associated with a vector multiplet
has a group theoretic interpretation. In 
particular the Born-Infeld field can be thought as
generating  an infinitesimal spinor rotation 
induced by a group $G$, where $G$ is again a
special holonomy group. 
Next we shall examine solutions of the 
Dirac-Born-Infeld action. Such solutions involve
non-vanishing vectors or tensors field 
as well as scalars. Although the supersymmetry
condition for such solutions can also be 
expressed in terms of spinor rotations, there is no
straightfoward interpretation of these 
conditions in terms of calibrations. We shall
summarize the solutions found which 
preserve $1/4$ of the supersymmetry and we shall present
some new ones which preserve $1/8$ of the supersymmetry.

This paper is organized as follows: In 
section two, we give the definition of calibrations
and summarize some of their properties. 
In section three, we show that all calibrations
preserve a proportion of the spacetime 
supersymmetry. In section four, we explain the
correspondence between worldvolume 
solutions and those of the bulk. In sections 
five, six and
seven, we give the correspondence between the 
worldvolume solutions associated with
calibrations and those of the bulk. In addition, 
we present all the associated supersymmetry
projections. In section eight, we give some of 
the singular worldvolume solutions. In
section nine, we find the worldvolume solutions 
which have only Born-Infeld fields. In section
ten, we relate these solutions to those 
associated with calibrations. In section eleven,
we investigate the worldvolume solutions 
which include both vectors and scalars and give
a new solution which preserves $1/8$ of 
the supersymmetry. Finally in section twelve,
we present our conclusions.

\chapter{Calibrations}

\section {The bound}

To investigate the supersymmetry of 
certain Born-Infeld configurations and 
establish bounds for their energy, we 
shall need a few facts about
calibrations [\cali,\calib]. We shall consider 
calibrations in $\bE^n$ equipped with the 
standard Euclidean inner product. Let 
$G(p,\bE^n)$ be the  
grassmannian, i.e. the space of (oriented)
 p-dimensional subspaces of $\bE^n$. 
Given a p-dimensional
subspace
$\xi$ of $\bE^n$, $\xi\in G(p,\bE^n)$,  we can always find an 
orthonormal basis $\{e_1, \dots, e_n\}$ in
$\bE^n$ such that such that $\{e_1, \dots, e_p\}$ is a basis in $\xi$.
We denote the co-volume of $\xi$ by
$$
{\buildrel \rightarrow\over\xi}=e_1\wedge \dots\wedge e_p\ .
\eqn\none
$$
 
A p-form $\phi$ on a open subset $U$ of $\bE^n$ 
is a calibration of degree $p$ if
\item{(i)} $d\phi=0$
\item{(ii)} for every point $x\in U$, the 
form $\phi_x$ satisfies $\phi_x({\buildrel
\rightarrow\over \xi})\leq 1$ for {\it all} 
$\xi\in  G(p,\bE^n)$ and such that the \lq
contact set'
$$
G(\phi)=\{ \xi\in G(p, \bE^n): \phi({\buildrel \rightarrow\over \xi})=1\}
\eqn\ntwo
$$
is not empty.

One of the applications of calibrations is that they 
provide a bound for the volume
of p-dimensional submanifolds of $\bE^n$. Let $N$ 
be a p-dimensional submanifold
of $\bE^n$.  At every point $x\in N$, one can find an 
oriented orthonormal basis in $\bE^n$
such that $\{e_1, \dots, e_p\}$ is an oriented orthonormal
 basis of $T_xN$. The co-volume of $N$ at $x$ is
 ${\buildrel \rightarrow\over N}= e_1\wedge \dots
\wedge e_p$ and the volume form of $N$ at $x$ is  
$\mu_N=\alpha_1\wedge \dots\wedge \alpha_p$, where 
$\{\alpha_1, \dots, \alpha_p\}$ is
the dual basis to $\{e_1, \dots, e_p\}$. The fundamental 
theorem of calibrations states the
following:
\item{\bullet} Let $\phi$ be a calibration of degree 
p on $\bE^n$. The p-dimensional submanifold
$N$ for which
$$
\phi({\buildrel \rightarrow\over N})=1
\eqn\nthree
$$
is volume minimizing. We shall refer to such minimal submanifolds as
calibrated submanifolds, or a calibrations for short, of degree $p$.

To prove the above statement, we choose an open subset $U$ of 
$N$ with boundary $\partial U$ and assume
that there is another subspace $W$ of $\bE^n$ and an open 
set $V$ of $W$ with the same boundary
$\partial U=\partial V$. Using the Stokes' theorem, we have
$$
{\rm vol} (U)=\int_U \phi=\int_V \phi=
\int \phi({\buildrel \rightarrow\over V}) \mu_V\leq \int_V
\mu_V ={\rm vol} (V)\ .
\eqn\nfour
$$

Calibrations give a large class of bounds for
the volume of subspaces in
$\bE^n$.
  The \lq charge'
density associated with the bound is given by the calibration form.
If $X$ is the map from the \lq worldvolume', 
$\bE^p$, into $\bE^n$. The above bound
 can be re-expressed as
$$
\int d^pu \sqrt{\rm{det} (g_{\mu\nu})}\geq \int X^*\phi\ ,
\eqn\nfive
$$
where $g_{\mu\nu}$ is the induced metric
 on $\bE^p$ with respect the map $X$ and $\{u^1,\dots, u^p\}$
are local  coordinates of $\bE^p$.

The  tangent spaces of a p-dimensional submanifold $N$ of $\bE^n$, 
parallel transported to the
origin of $\bE^n$, span a subspace of $G(p,\bE^n)$; 
this is the Gauss map. If moreover $N$
saturates the bound associated with the
 calibration $\phi$, then the image of the Gauss
map is in $G(\phi)$.
In many cases $G(\phi)=G/H$, where $G$ is a subgroup of $SO(n)$.
As we shall see, the group $G$ 
arises naturally in the  investigation of the supersymmetry of
such configurations.  There are many
examples of calibrations. Here we shall 
mention three examples which will 
be proved useful in the study
of supersymmetry and in the construction of actual solutions of the 
field equations of the Nambu-Goto action.

There is a relation between calibrations and
special holonomy groups.  This is because special holonomy
groups are characterized by the existence of certain
 invariant forms.  These 
forms can be used as calibrations. 
Manifolds with special holonomy also 
admit covariantly constant spinors (see the  table below). 
Such spinors are also invariant
under the action of the holonomy groups; the 
invariant spinors and forms are
related [\wang]. 
\vskip 0.5cm
$$
\vbox{\settabs 4\columns 
\+ {\rm Holonomy}&Forms& Spinors& Dim \cr
\+ $SU(n)$& 2(1), n(1)&~~ 2&2n\cr
\+ $Spin(7)$&$4^+(1)$&~~~1&8\cr
\+ $G_2$&3(1),4(1)&~~~1&7\cr
\+ $Sp(n)$&2(3)&~~n+1&4n\cr}
$$
\vskip 0.2cm
\noindent{\bf Table 1: Covariantly Constant Forms 
and Spinors} \ \ The first column contains
the special holonomy groups, the second 
column contains the degrees
(multiplicities) of covariantly constant 
forms ($4^+$ denotes a self-dual four-form), the
third column contains the dimension of the 
space of covariantly constant spinors and the
last column contains the dimension of the 
special holonomy manifold. 
\noindent
\vskip 0.3cm
The investigation
 of supersymmetric
solutions of  Nambu-Goto action provides a further
 connection between calibrations
and special holonomy groups. This connection 
makes use of the invariant spinors
of special holonomy groups. In particular, 
the spinor representations\foot{Our
metric convention is $\eta={{\rm diag}( 1, \dots, 1, -1)}$} of 
${\rm Cliff}(9,1)$ or ${\rm Cliff}(10,1)$ have
 singlets when decomposed under the special
holonomy subgroups of ${\rm Spin}(9,1)$ or
${\rm Spin}(10,1)$.
As we shall show, if a calibration as solution 
of the Nambu-Goto action is associated with a
special holonomy group, then
 the invariant spinors serve as Killing spinors 
and some of the spacetime supersymmetry is
preserved.  The proportion of
the supersymmetry preserved is related to the  
number of singlets in the decomposition of the 
spinor representations of
${\rm Spin}(9,1)$ under the special holonomy subgroup.
We shall remark further on this relationship between special holonomy,
calibrations and superymmetry in the section eleven.

\section{K\"ahler Calibrations}

We begin by introducting coordinates $\{x^i, y^i;i=1,\dots, n\}$ 
 on $\bE^{2n}$ and the metric
$$
ds^2=\sum^n_i( dx^i dx^i+dy^i dy^i)\ .
\eqn\kone
$$ 
Then we choose a complex structure on $\bE^{2n}$ 
such that the associated K\"ahler form is
$$
\omega={i\over2} \sum_i dz^i\wedge d\bar z^i
\eqn\mone
$$
where $z^i=x^i+iy^i$. Next we set
$$
\phi={1\over p!} \omega^p\ .
\eqn\ktwo
$$ 
The form $\phi$ is a calibration of degree $2p$ and the
 contact set $G(\phi)$ is the space of complex
p-dimensional planes in $\bC^n=\bE^{2n}$,
 $G(\phi)=G_{\bC}(p, \bC^n)$, where we have identified $\bE^{2n}$
with $\bC^n$ using the above complex structure. To prove this, one uses 
Wirtinger's inequality which states that
$$
\phi({\buildrel \rightarrow\over\xi})\leq 1
\eqn\mthree
$$
for every $\xi\in G(2p, \bE^{2n})$ with equality 
if and only if $\xi\in G_{\bC}(p, \bC^n)$ (for
the proof see [\cali, \calib]). 
A consequence of this is that all complex 
submanifolds of $\bC^n$ are volume minimizing. The 
form $\omega$, and therefore $\phi$, are
invariant under $U(n)$. The contact set can be
 written as the coset space 
$$
G_{\bC}(p, \bC^n)=U(n)/U(p)\times U(n-p)=SU(n)/S(U(p)\times U(n-p))\ .
\eqn\mfour
$$
Thus $SU(n)$ acts transitively on the 
space of p-dimensional complex
planes in $\bC^n$. As we shall see this 
fact will be used to find
the proportion of spacetime supersymmetry
 preserved by the worldvolume solutions associated
with K\"ahler calibrations.

\section{Special Lagrangian Calibrations}

To describe the special Lagrangian 
calibrations we begin with the  metric \kone\ and the
 K\"ahler form $\omega$ \ktwo\ as in the previous section.
In addition, we introduce the $(n,0)$-form
$$
\psi=dz^1\wedge \dots\wedge dz^n\ .
\eqn\slone
$$
 The data, which include the metric, the K\" ahler form and 
the $(n,0)$-form $\psi$, are invariant under
$SU(n)$.  The 
calibration form\foot{We can define special Lagrangian calibrations with a more general
$(n,0)$-form. However for this paper this choice will suffice.} is  in
this case is
$$
\phi={\rm Re}\psi\ .
\eqn\sltwo
$$
The inequality necessary for $\phi$ to be a calibration
 has been demonstrated in [\cali, \calib] and the
planes that saturate the bound are called special 
Lagrangian.  The contact
set in this case is 
$$
G(\phi)=SU(n)/SO(n)\ .
\eqn\slthree
$$
As in the case of K\"ahler calibrations, $SU(n)$ acts 
transitively on the space of special
Lagrangian planes.

\section{Exceptional Cases}

There are three exceptional cases. 
\item{(i)} The
calibration form is the 3-form, $\varphi$, in $\bE^7$ 
invariant under the exceptional
group  $G_2$.  This gives rise to a calibration 
of degree three in $\bE^7$. The 
contact set is 
$$
G(\varphi)=G_2/SO(4)
\eqn\slfour
$$
which is a subset $G(3, \bE^7)$. 
\item{(ii)} The calibration 
form is the dual $\chi$
of
$\varphi$ in $\bE^7$, $\chi=\star \varphi$.  This gives rise to
 a calibration of degree
four in $\bE^7$. The contact set is again 
$$
G(\chi)=G_2/SO(4)
\eqn\eone
$$
which is now thought of as the subset of $G(4, \bE^7)$. 
\item{(iii)}  The calibration
 form is the ${\rm Spin}(7)$-invariant self-dual
4-form, $\Phi$, in $\bE^8$.  This gives rise 
to a calibration of degree four
in $\bE^8$.  The contact set is
$$
G(\Phi)= {\rm Spin}(7)/H
\eqn\ethree
$$
which is a subset of $G(4, \bE^8)$, where 
$H=SU(2)\times SU(2)\times SU(2)/\bZ_2$. 
This calibration is
called a Cayley calibration and the 4-planes that 
saturate the bound are called Cayley planes.

\chapter{ Supersymmetry and calibrations}

The dynamics of a large class of extended objects is described by  
actions of Dirac-Born-Infeld type. The fields
include the embedding maps $X$ of the extended object into 
spacetime which we shall take it to be 
ten-dimensional Minkowski, $\bE^{(9,1)}$. Apart from
the embeddings maps, the action  may depend on 
other worldvolume fields like a Born-infeld
2-form field strength
$F$ (as in the case of D-branes)  and the various fermionic partners
$\theta$ which are spacetime fermions. These actions
are invariant under fermionic transformations which are  
commonly  called kappa-symmetries. These 
 act on $\theta$ as
$$
\delta\theta=(1+\tilde\Gamma)\kappa
\eqn\pone
$$
where $\tilde\Gamma\in {\rm Cliff}(9,1)$  is a
traceless hermitian product  structure, {\sl i.e.} 
$\tilde\Gamma^2=1$, and $\kappa$ is the
parameter which is a spacetime fermion. The 
product structure $\tilde\Gamma$ 
depends on the embeddings maps and the other
worldvolume fields, {\sl i.e.} $\tilde\Gamma$ depends on the
 worldvolume coordinates. It was shown in
[\kallosh] that the Killing spinor equation in this
 context is the algebraic equation
$$
(1-\tilde\Gamma)\epsilon=0\ ,
\eqn\susyproj
$$
where $\epsilon$ is the spacetime supersymmetry parameter.  
Since we have chosen the spacetime
geometry to be Minkowski, the spacetime (supergravity)
 Killing spinor equations imply that
$\epsilon$ is constant\foot{If we relax the condition 
that $\epsilon$ is constant, then the
Killing spinor equation \susyproj\ has always solutions.}. 
The condition for supersymmetry
\susyproj\ is universal and applies to all 
types of branes, fundamental strings, 
solitonic 5-branes,
D-branes and M-branes. {\sl Supersymmetric 
(worldvolume) configurations are solutions
of the Born-Infeld field equations which 
 satisfy \susyproj\ for some non-vanishing
$\epsilon$}. The proportion of the bulk 
supersymmetry preserved by such configuration depends
on the number of linearly independent 
solutions of \susyproj\  in terms of $\epsilon$. 

The  product structure $\Gamma$ for 
D-branes and the M-5-brane can be written [\kallosh] as
$$
\tilde\Gamma=e^{-{a\over2}}  \Gamma e^{{a\over2}}
\eqn\ptwo
$$
where $\Gamma$ depends only on the embedding
map of the worldvolume into
spacetime and $a$ contains the depence of
 $\tilde \Gamma$ on the Born-Infeld
fields. 
If we take the Born-Infeld fields to vanish, 
then $a=0$ and the product structure
becomes
$$
\tilde\Gamma=\Gamma \ . 
\eqn\pthree
$$

\section{Supersymmetry without Born-Infeld fields}

A consistent truncation of all Dirac-Born-Infeld 
type of actions is to allow the Born-Infeld
fields to vanish. Let $X$ be the embedding map 
of the extended object into Minkowski
spacetime. The 
 bosonic part of the
truncated action is 
$$
S_{NG}=\int d^{p+1}x \sqrt {|det (g_{\mu\nu})|}
\eqn\pfour
$$
where 
$$
g_{\mu\nu}=\partial_\mu X^M \partial_\nu X^N \eta_{MN}\ ,
\eqn\pfive
$$
is the induced metric on the worldvolume with 
coordinates $\{x^\mu; \mu=0,\dots, p\}$,
and $\eta$ is the ten-dimensional Minkowski metric.
The product structure $\Gamma$ is then expressed 
in terms of the embedding maps
and the spacetime Gamma matrices.  The 
particular expression for $\Gamma$ depends
on the type of brane that we are considering.  
Here we shall investigate the 
Killing spinor equations for IIA D-branes. However 
our argument is universal and applies
equally well to IIB D-branes, M-branes, 
fundamental strings and heterotic 5-branes.

To continue, we define
$$
\gamma_\mu=\partial_\mu X^M \Gamma_M
\eqn\psix
$$
where $\{\Gamma_M; M=0,\dots,9\}$ are the 
spacetime gamma matrices. We remark that
$$
\gamma_\mu \gamma_\nu+\gamma_\nu \gamma_\mu=2 g_{\mu\nu}
\eqn\pseven
$$
{\sl i.e.} the $\{\gamma_\mu ; \mu=0,\dots, p\}$ obey 
the Clifford algebra with respect
to the induced metric. The product structure $\Gamma$ for IIA D-branes is
$$
\Gamma={1\over (p+1)! \sqrt g} \epsilon^{\mu_0\dots \mu_{p}} 
\gamma_{\mu_0}\dots \gamma_{\mu_{p}} \Gamma^p_{11}\ .
\eqn\peight
$$
For example, the product structure of a planar 
IIA D-p-brane which spans the first
$p$ directions of $\bE^{(9,1)}$ is
$$
\Gamma=\Gamma_{0}\dots \Gamma_p \Gamma^p_{11}\ .
\eqn\pnine
$$

There is another way to describe the 
product structure $\Gamma$. For this we remark
that there is a map $c$ which assigns to every q-form
$$
\nu={1\over q!} \nu_{M_1,\dots M_p} dx^{M_1}\wedge\dots\wedge dx^{M_p} 
\eqn\wone
$$
 in $\bE^{(9,1)}$ an element
$$
c(\nu)={1\over q!} \nu_{M_1,\dots M_p} \Gamma^{M_1}\dots \Gamma^{M_p}
\eqn\wtwo
$$ 
of the Clifford algebra ${\rm Cliff}(9,1)$.
Note that 
$$
c(\star\nu)= c(\nu)\Gamma_{11}
\eqn\wthree
$$
where $\star\nu$ is the Hodge dual of $\nu$.
 Let $N$ be the submanifold of $\bE^{(9,1)}$
that describes a solution of the Born-Infeld 
field equations. Then the product structure
$\Gamma$ associated with $N$ is
$$
\Gamma=c(\mu) \Gamma^p_{11}
\eqn\planar
$$
where $\mu$ is the volume form of $N$.
Therefore, the product structure $\Gamma$ 
written in terms of an oriented orthonormal
frame at a point $y$ becomes the product 
structure of a planar p-brane tangent to $N$ 
at $y$. {\sl Thus to solve the Killing
spinor equation of $N$ is equivalent to  
requiring that the Killing spinor equations of the
planar D-p-branes  tangent to $N$ at each
 point $y$ have a common solution.}

One way to find a solution is to recall
 what happens when two planar 
branes intersect at an \lq\lq angle". 
Supersymmetry requires that
the element of $SO(9,1)$ or $SO(10,1)$ 
relating the hyper-planes is in a subgroup 
$G\subset SO(9,1)$ or $SO(10,1)$. Such 
angles are called \lq\lq $G$-angles". For a general
non-planar brane we proceed analogously. 
 For this we choose a reference point $y_0$ and an
orthonormal frame of the tangent space
$T_{y_0}N$  at this point which we extend to 
an orthonormal  frame in $\bE^{(9,1)}$.  Then we
choose any other point
$y$ in $N$ and introduce another orthonormal 
frame in the same way. The two orthonormal
frames are related by a Lorentz transformation
 in $\bE^{(9,1)}$ (a spatial orthogonal
 rotation for static configurations.). Moreover 
the supersymmetry condition
at $y$ can be written in terms of the product 
structure at $y_0$ as
$$
S^{-1} \Gamma(\{y_0\}) S\epsilon=\epsilon
\eqn\rotproj
$$
where $S$ is a spinor rotation induced by the
 Lorentz rotation that relates
the above orthonormal frames at $y_0$ and $y$. 
For most of the
solutions of the Dirac-Born-Infeld equation, the rotations required
are generic elements of the ten-dimensional 
Lorentz group and all
supersymmetry is broken.  However, 
some cases involve rotations which lie in a 
subgroup of the Lorentz group.
In particular, if the rotations lie in a 
subgroup of the Lorentz group
for which the decomposition of the  Majorana
 spinor representation of ${\rm Spin}(9,1)$
 has singlets, then the equation \rotproj\ reduces to
$$
\Gamma(\{y_0\}) \epsilon=\epsilon
\eqn\wfour
$$
provided that $\epsilon$ is a linear 
combination of the singlets. Therefore
the Killing spinor equation of $N$ in such a 
case reduces to the Killing
spinor equation on a single D-p-brane which can
 be easily solved. Examples of subgroups of the
orthogonal groups
 for which the decomposition of
the ten-dimensional spinor representation
representation has singlets are  the special
holonomy groups like $SU(k)$, $Sp(k)$,
$G_2$ and ${\rm Spin}(7)$.

It is clear from the above arguments that 
preservation of a 
proportion of supersymmetry of the bulk
by a Dirac-Born-Infeld configuration is 
closely related to calibrations.  Let $\Phi$
be the Gauss map which takes the tangent space of 
$N$ at every point $y$ into
the grassmannian $G(p+1, \bE^{(9,1)})$. If the image 

$$
S_{\Gamma}=\Phi(N)
\eqn\wfive
$$ 
of such a map in
$G(p+1, \bE^{(9,1)})$ is a subspace of the 
homogeneous space $G/H$, then 
clearly the relevant
rotations
 amongst the frames associated with the 
tangent spaces are in $G$. If $N$
is a calibration, then $S_{\Gamma}$ is a 
subspace of the contact set $G(\phi)$. In all the
examples of calibrations that we have presented, 
the contact set is a homogeneous space with
$G$ being one of the groups that appear in the 
context of special holonomies. From this
we conclude that all these calibrations preserve
 a proportion of the supersymmetry of the
bulk.  Clearly the case of intersecting planar
 branes is a special case
of this construction. For more about 
this see section eight.

\section{ Supersymmetry with Born-Infeld fields}

The Dirac-Born-Infeld action with Born-Infeld field $F_{\mu\nu}$ is
$$
I=\int d^{p+1}x {\sqrt{|{\rm det}\big(g_{\mu\nu}+F_{\mu\nu}\big)|}}
\eqn\fone
$$
where $g_{\mu\nu}$ is induced metric. This  action
describes the dynamics of D-p-branes as well as the dynamics of
IIB NS-NS 5-branes.  Here we shall investigate the 
Killing spinor equations for IIA D-branes. However
 our argument can be easily
extended to  apply in the case of IIB D-branes, 
in the case IIB NS-NS 5-brane 
as well as in the case of
M-5-brane.
For IIA D-p-brane [\kallosh],
$$
a=Y_{ \mu \nu} \gamma^{\mu} \gamma^{ \nu}\Gamma_{11}\ ,
\eqn\ftwo 
$$
where
$$
F=\lq\lq {\rm tan}" Y\ .
\eqn\tana
$$

The inclusion of non-vanishing Born-Infeld
 fields modifies the conditions
for a configuration to be supersymmetric. 
In this case, there is no a direct relation between
calibrations and supersymmetry conditions. 
Nevertheless the analysis for the proportion of supersymmetry
 preserved by a
solution of the Dirac-Born-Infeld field equations
proceeds as in the previous case without 
Born-Infeld fields. We begin with a submanifold $N$
in $\bE^{(9,1)}$ which is a solution of the 
Born-Infeld field equations together
with a non-vanishing Born-Infeld field $F$. Then 
we introduce orthonormal frames at the
tangent spaces $T_{y_0}N$ and $T_{y_1}N$ of the points
points  $y_0$ and
$y_1$ of $N$ which we extend to orthonormal frames
 in $\bE^{(9,1)}$. Again there is a
 Lorentz rotation in  $\bE^{(9,1)}$ which 
relates the two frames. Using a similar argument
to that of the previous section, the Killing
 spinor equation at a point $y$
can be written in terms of the Killing spinor
 equation at $y_0$ as follows
$$
U^{-1} \big(e^{-{a(y_0)\over 2}} \Gamma(y_0) e^{a(y_0)\over 2}\big)
 U\epsilon=\epsilon  
\eqn\fthree
$$
where 
$$
U=e^{-{a(y_0)\over 2}} S e^{ a(y)\over 2}
\eqn\ffour
$$
and 
$S$ is the induce rotation on the spinors
 from the Lorentz rotation that relates
the two frames.
These supersymmetry projections can be
simplified if we assume that the
Born-Infeld field vanishes at the
 reference point $y_0$.  (In many applications, the
Born-Infeld field vanishes at some point
 usually at infinity.) The supersymmetry projection at
$y$ then becomes
$$
e^{-{ a(y)\over 2}} S^{-1}   \Gamma(y_0) 
  S e^{ a(y)\over 2}\epsilon=\epsilon\ . 
\eqn\ffive 
$$
Solutions of these conditions can be found
by assuming that the spinor-rotations $U$ leave
invariant some {\sl constant} 
Majorana spinor $\epsilon$. In such a case,
the Killing spinor equation for $N$ 
degenerates to the Killing spinor equation of the
reference point
$$
\Gamma(y_0)\epsilon=\epsilon\ ,
\eqn\fsix
$$
where $\epsilon$ is a linear
 combination of the singlets.

A consistent truncation of the Dirac-Born-Infeld
 action is to set $\{X^M\}=\{x^\mu, y^i=0\}$.
The truncated 
action is the Born-Infeld action
$$
I=\int d^{p+1}x {\sqrt{{\rm det}\big(\eta_{\mu\nu}+F_{\mu\nu}\big)}}\ .
\eqn\fseven
$$ 
For this class of configurations, $N$ is 
a (p+1)-dimensional Minkowski subspace
of $\bE^{(9,1)}$. Therefore we can choose
 an orthonormal frame in $\bE^{(9,1)}$
such that $N$ spans the first $p$ spatial 
directions of $\bE^{(9,1)}$.
The Killing spinor equations become
$$
e^{-{a(y)\over 2}} \Gamma e^{a(y)\over 2}\epsilon=\epsilon
\eqn\feight
$$
where $\Gamma$ is a constant product structure.

In the linearized limit, this condition on 
the Born-Infeld field $F$ reduces to the
familiar condition for a (Yang-Mills)
 configuration to preserve a proportion of
supersymmetry\foot{We have assumed that $F$ 
is not a constant two-form field strength.}
together with a projection of associated with
$\Gamma$.  Viewing
$F$ as the infinitesimal transformation of
$SO(p,1)$ rotations acting on the Killing spinor
 with the induced spinor representation,
supersymmetry is preserved provided that
$F$ takes values in an appropriate subalgebra of 
$so(p,1)$. Such subalgebras are those of
the special holonomy subgroups of $so(p,1)$.  For example
$sp(1)$ is related to the self-dual 
 connections in four dimensions,
$su(k)$ is related to the Einstein-Yang-Mills
 connections [\donald], $sp(k)$ is related to the
instanton solutions found in [\kent] 
(see also [\gptf]) and similarly for $g_2$ and $spin(7)$
[\fn, \ivanova, \gn].

\chapter{Static Born-Infeld Solitons, Brane Boundaries
 and Brane Intersections }

As we have already mentioned, the worldvolume 
solitons of the Dirac-Born-Infeld action of a
p-brane may arise as the intersection of the
 p-brane with other branes or  the boundary of
other branes ending on the p-brane. Therefore, 
such worldvolume solitons can be found
by utilizing the bulk boundary and intersection 
rules for branes. These rules are known either
from string theory or from supegravity.  
Worldvolume solitons that arise as boundaries or
 intersections of branes have the following
qualitative properties:
\item{\bullet} The bulk brane configuration and 
the associated worldvolume
soliton exhibit the same manifest  Poincar\'e invariance.
\item {\bullet} The proportion of the 
bulk supersymmetry preserved by the 
worldvolume soliton is the same as that of 
the associated bulk configuration.
We shall show that supersymmetric projections
 in both cases are identical. 
\item{\bullet} The worldvolume solitons of a
 p-brane that are associated with other branes
ending on it have non vanishing Born-Infeld fields.  
This is a consequence of the
conservation of the flux at the intersection.
\item{\bullet} The number of non-vanishing
 scalar fields associated with
a worldvolume soliton of a p-brane is equal 
to the number of the worldvolume directions
of the other branes involved in the associated 
bulk configuration which are transverse to
 the p-brane.
 
For worldvolume solitons associated with the 
intersection of two branes, we can
also impose the following boundary conditions:
\item {\bullet}  Away from the boundary or the intersection, the 
induced metric on the p-brane should
approach that of $\bE^{(p,1)}$.
\item {\bullet} Near the boundary or the 
intersection on a p-brane 
the induced metric should approach
the worldvolume Minkowski metric of the
 \lq incoming' brane from the bulk.

There is a very large number of brane 
configurations that have the interpretation
of branes ending on or intersecting with other branes.  We shall not attempt to  
present a complete list here. However, 
we shall mention some examples below (see [\gppkt,
\tseytlinb, \kastor]). These bulk 
configurations are most easily classified according to the
supersymmetry that they preserve.
\item{(i)} Configurations that preserve 
1/2 of the supersymmetry of the bulk are bound states
of branes that lie within other branes. These bound states are below threshold.
Examples of such bound states include 
the D-(p-2)-branes within D-p-branes and the
M-2-brane within the M-5-brane.

\item{(ii)}Configurations that preserve 
1/4 of the supersymmetry of the bulk include the
following: In M-theory, two membranes
 intersecting at a 0-brane, two M-5-branes 
intersecting at a
3-brane, a membrane ending on a 5-brane 
with boundary a string.  In string theory we have,
the fundamental string ending on a D-p-brane
 with a 0-brane boundary, a D-string ending at a IIB
NS-NS 5-brane with a 0-brane boundary, two
 D-p-branes intersecting at a D-(p-2)-brane and two
NS 5-branes intersecting at a 3-brane.

\item {(iii)} Configurations that preserve 1/8 of the 
supersymmetry of the bulk  include
the following: In M-theory, three M-2-branes 
intersecting at a 0-brane, three M-5-branes
intersecting at a string and two M-2-branes 
ending a M-5-brane with their string boundaries
intersecting on a 0-brane.

Using the correspondence between bulk
 configurations and worldvolume solitons, it
 is worth mentioning that
there are two 0-brane worldvolume solitons on the D-2-brane. 
 One is due to the IIA fundamental
string ending on the D-2-brane and the other is due to the 
intersection of two D-2-branes at a
0-brane.  The two 0-brane solitons are 
charged with respect to different fields.
The first one is charged with respect 
to the Born-Infeld one-form gauge potential and the
other is charged with respect to 
the gauge potential associated with the dual of a transverse
scalar. The D-4-brane also has two 
different 0-brane worldvolume solitions. One is
due to the fundamental string ending 
to the D-4-brane and it charged with respect to
the Born-Infeld gauge potential. The other is 
due to a D-0-brane within the D-4-brane and the
worldvolume solution is a Born-Infeld field instanton.

\chapter{Static Solutions and K\"ahler calibrations}

The worldvolume solitons associated with
 K\"ahler calibrations are complex submanifolds
of $\bC^n$.
If the calibration has contact set
$SU(n)/S(U(p)\times U(n-p))$, then a generic
 soliton  will preserve
$1/2^{n}$ of the supersymmery of the bulk. 
As we shall see, the singlets of the
ten-dimensional representations under 
$SU(n)\subset SO(9,1)$ satisfy $n-1$ relations
each reducing the number of components 
of the Killing spinor by half. The 
supersymmetry projection associated 
with the p-brane also reduces the components
of the Killing spinor by another
 half leading to the preservation of the $1/2^n$
of the bulk supersymmetry.

The worldvolume solitons associated 
with the K\"ahler calibrations have vanishing
Born-Infeld type of fields and 
therefore the analysis below applies 
universally to all types
of branes, M-branes, D-branes and NS-branes.   Let us 
describe the solution of a k-brane soliton of a p-brane with
$\ell$ transverse scalars.  For this 
we choose the static gauge for the p-brane,
$$
X^M=(x^\mu, y^i)\ ,
\eqn\staticgauge
$$
where $\{\mu=0, \dots , p\}$ are the worldvolume coordinates 
and $\{y^i;i=1, \dots, 9-p\}$ are the
transverse coordinates of the p-brane. 
In this gauge,  the Born-Infeld
solution that we would like to describe 
is a map $y$ from $\bE^{p-k}\subset \bE^{(p,1)}$ into
$\bE^\ell\subset \bE^{(9,1)}$.  We then 
introduce  complex structures $I$ and $J$ on
$\bE^{p-k}$ and 
$\bE^\ell$, respectively, and we write the equation
$$
I^b{}_a\partial_b y^m=J^m{}_n \partial_ay^n\ ,
\eqn\kbog
$$
where $a,b=1, \dots, p-k$ and $m,n=1, \dots, \ell$; the dimension 
of $\bE^{p-k}$ and $\bE^\ell$ are
even.   It is straightforward to show that if $y$ 
satisfies \kbog\ then the embedding
$\{X^M\}=\{(x^\mu, y^m,0)\}$ solves the Born-Infeld field equations. 
 Choosing complex coordinates with respect
to these complex structures, $x^a=(z^\alpha, \bar z^{\bar \alpha})$ 
and $y^m=(s^A, \bar s^{\bar A})$, the
solutions of \kbog\ are  holomorphic functions 
$$
s^A=s^A(z)\ .
\eqn\fnine
$$
Therefore the above solution is a K\"ahler 
calibration in $\bE^{n-k+\ell}$ of
degree
$p-k$.  The K\"ahler form is given by the 
complex structure $I\oplus J$ and the contact set
is
$$
SU({p-k+\ell\over2})/S(U({p-k\over 2})\times U({\ell\over2}))\ .
\eqn\ften
$$

\section{SU(2) K\"ahler calibrations}

The simplest of all K\"ahler calibrations is 
the one describing an one-dimensional
complex space in $\bC^2$.  The associated 
Born-Infeld brane solitons have already being found
in [\gary]. They correspond to (p-2)-brane 
worldvolume solitons that appear in all p-branes
($p\geq 2$). From the bulk (supergravity) 
perspective, the existence of these solitons is due 
 to the
\lq\lq (p-2)-intersection rule" which states 
that two p-branes intersect on a (p-2)-brane
[\gppkt].  The solution can be written  in an implicit
form as
$$
F(s,z)=0\ .
\eqn\zone
$$ 
Different choices of $F$ give different solutions. Some  examples are the
following. (i) The 3-brane soliton of M-5-brane 
for $F$ the equation of a 
Riemann surface of was investigated in [\witten]. (ii) In [\krogh],
the 0-brane soliton of the M-2-brane for some 
choice of $F$ was interpreted as
the M-theory analogue of the IIB 
(p,q)-string triple junctions [\schwarzb, \sena] (for more references to earlier
work see [\rey]).  
Another example is to choose
$$
F(s,z)=s z-c\ ,
\eqn\ztwo
$$
where $c$ is a complex number.
 If $c=0$, then the
soliton is singular. The bulk interpretation of
 such a soliton is that of two planar p-branes
intersecting at a (p-2)-brane with the singularity 
located at the intersection. Such
solutions will be investigated in section 8.  If
$c\not=0$, the singularity is blown up and the 
induced metric on the p-brane for this
configuration has two asyptotically flat regions 
one at $|z|\rightarrow \infty$ and the 
other at $|z|\rightarrow 0$.  These asymptotic regions are
identified with the two p-branes involved in the 
intersection. It turns out that the
two asymptotic regions are orthogonal in the bulk metric 
and therefore the intersection is at
right angles. Next, we take
$$
F(s,z)=(s-bz) z-c\ ,
\eqn\zthree
$$
where $b,c$ are complex numbers.
The induced metric on the p-brane again has two asymptotically flat 
regions one at $|z|\rightarrow \infty$ and the 
other at $|z|\rightarrow 0$ which can be 
identified with the p-branes involved
in the intersection. However in this case the
 intersection is not orthogonal. This
can be easily seen by observing that 
$|s|\rightarrow \infty$ as $|z|\rightarrow 0$
and $s\rightarrow bz$ as $|z|\rightarrow \infty$.  
Therefore in the $(s,z)$ plane, one
asymptotic region is along the $s$ plane while
 the other is along the plane determined
by the equation $s=b z$. The two angles among 
these two planes are determined by the
parameter $b$; for $b$ a real number, 
$$
\tan \theta={1\over b}
\eqn\zfour
$$ 
where $\theta$ is the the angle.

As we have already mentioned all the above
 configurations preserve $1/4$ of the
bulk supersymmetry. To be more explicit 
let us consider the case
of IIA D-branes in more detail. We can 
always arrange, without loss of generality, 
that the product structure $\Gamma$ at the reference point $y_0$ is
$$
\Gamma(y_0)=\Gamma_0\dots \Gamma_p (\Gamma_{11})^{{p+2\over2}}\ .
\eqn\zfive
$$
We shall consider solutions which depend on 
the worldvolume coordinates
$(x^{p-1}, x^p)$$=(X^{p-1}, X^p)$ and with 
 transverse coordinates $(y^1,
y^2)=$$(X^{p+1}, X^{p+2})$. We then introduce 
the complex coordiantes $z=x^{p-1}+i x^p$ and
$s=y^1+i y^2$. 
The supersymmetry projection associated with $y_0$ is
$$
\Gamma_0\dots \Gamma_p (\Gamma_{11})^{{p+2\over2}}\epsilon=\epsilon\ .
\eqn\susyoneii
$$
For such a K\"ahler calibration, the
$SU(2)$ rotations  of the tangent
 planes of the solution is
taking place in the space spanned by 
the coordinates $\{X^{p-1}, \dots, X^{p+2}\}$. For
the above choice of complex structure,
the spinor singlets of any $SU(2)$ rotation
in these directions satisfy
$$
\Gamma_{p-1}\Gamma_p\Gamma_{p+1} \Gamma_{p+2}\epsilon=-\epsilon\ ;
\eqn\susytwoii
$$
(the sign depends on the choice of complex 
structure). Combining the conditions
\susyoneii\ and \susytwoii, we conclude 
that the supersymmetry preserved by the worldvolume
solution is
$1/4$ of that of the bulk.
It is worth pointing out that the two
 supersymmetry conditions can be rewritten as
$$
\eqalign{
\Gamma_0\dots \Gamma_p (\Gamma_{11})^{{p+2\over2}}\epsilon&=\epsilon
\cr
\Gamma_0\dots \Gamma_{p-2} \Gamma_{p+1}\Gamma_{p+2}
(\Gamma_{11})^{{p+2\over2}}\epsilon&=\epsilon\ .}
\eqn\xone
$$
These are precisely the supersymmetry conditions
 of the supergravity solution with the
interpretation of two IIA D-p-branes intersecting 
on a (p-2)-brane\foot{The intersection
can be at any $SU(2)$ angle.}. 
This turns out to be a common feature of all 
supersymmetric worldvolume solitons. 
If the worldvolume soliton has a bulk 
interpretation, then {\sl the supersymmetry
conditions of the soliton are identical to 
those of the bulk configuration}.
The above argument for supersymmetry with
a slight modification applies
to all (p-2)-brane worldvolume solitons of p-branes.

\section {SU(3) K\"ahler calibrations}

There are two cases of K\"ahler calibrations
 associated with the group $SU(3)$
to consider.  The first case is a degree 
two calibration in $\bC^3$ and the second case 
is a degree four calibration in $\bC^3$. Both cases have the same contact set.

The worldvolume solitons of a p-brane 
associated with the calibration of degree two 
are described be the zero locus of the holomorphic functions
$$
\eqalign{
F^1(s^1, s^2, z)&=0
\cr
F^2(s^1, s^2, z)&=0\ ,}
\eqn\xtwo
$$
where $z$ is a complex coordinate on the 
p-brane, $p\geq2$, and $s^1, s^2$ are complex
coordinates transverse to the p-brane.   
Next suppose that 
$z=x^{p-1}+ix^p$, $s^1=y^1+iy^2$ and 
$s^2=y^3+iy^4$ ($y^i=X^{p+i}$).  For this choice of
complex structure, the spinor singlets
 of $SU(3)$ acting on $z, s^1, s^2$  satisfy
the conditions
$$
\eqalign{
\Gamma_{p-1}\Gamma_p\Gamma_{p+1} \Gamma_{p+2}\epsilon&=-\epsilon
\cr
\Gamma_{p-1}\Gamma_p\Gamma_{p+3} \Gamma_{p+4}\epsilon&=-\epsilon\ .}
\eqn\susyq
$$
We then choose  the base point
$y_0$ as in the $SU(2)$ case above so that 
the projector of the p-brane is given as
in \susyoneii. Therefore a generic
 worldvolume soliton preserves $1/8$  of the
bulk supersymmetry as expected.

From the bulk perspective, the above 
worldvolume solitons are associated
with the common intersection of 
three p-branes. There are several
examples of such configurations.  The 
typical example is that of three M-2-branes (or
equivalently three D-2-branes) 
intersecting on a 0-brane. The intersection that
we are considering is not
necessarily orthogonal. However to simplify notation, 
we shall present here
the orthogonal bulk intersection which is 
$$
\eqalign{
{\rm (i)\, M-2:}&\qquad 0,\,1,\,2,\,-,\,-,\,-,\,-
\cr
{\rm (ii)\, M-2:}&\qquad 0,\,-,\,-,\,3,\,4,\,-,\,-
\cr
{\rm (iii)\, M-2:}&\qquad 0,\,-,\,-,\,-,\,-,\,5,\,6\ .}
\eqn\inttwo
$$
In this notation, the numbers denote the bulk
 directions which are identified with the
worldvolume directions of the associated brane 
 and some of the transverse  directions
are denoted
$-$.
 From the perspective of one of the three 2-branes, 
say the first one, $z$ spans the
worldvolume directions $1,2$ and  $s^1, s^2$ span the
worldvolume directions $3,4,5,6$ of the other two 2-branes.  
Using reduction from M-theory to IIA and T-duality, 
we can construct many other cases
like for example that of three D-3-branes 
intersecting at a string. 

An example of a worldvolume soliton that
 corresponds to the above intersection is
$$
\eqalign{
F^1&=(s^1 -b^1 z) z-c^1
\cr
F^2&=(s^2 -b^2 z) z-c^2\ ,}
\eqn\xxone
$$
where $b^1,b^2,c^1,c^2$ are complex numbers. The 
induced metric on one of the M-2-branes
has two asymptotically flat regions, one as 
$|z|\rightarrow 0$ and the other as
$|z|\rightarrow
\infty$. Identifying $s^1,s^2$ with the
 coordinates of the \lq incoming' branes,
comparing the asympotic behaviour of the
 solution as $|z|\rightarrow 0$ and $|z|\rightarrow
\infty$ and using a similar argument to that 
of the $SU(2)$ case, we find that the angles
are determined by $b^1$ and $b^2$; for $b^1,b^2$ real numbers, we have
$$
\eqalign{
\tan\theta^1&={1\over b^1}
\cr
\tan\theta^1&={1\over b^2}\ .}
\eqn\xxthree
$$
If  the constants $\{b^1, b^2\}$
vanish, then the branes intersect orthogonally. 
It is worth mentioning that the
supersymmetry projections associated with 
this worldvolume soliton are the same as
 the supersymmetry projections associated 
with the bulk configuration \inttwo.

Next we shall consider the case of degree
 four calibrations in $\bC^3$.  The worldvolume
solitons of a p-brane are described by the 
zero locus of the holomorphic function
$$
F(z^1, z^2, s)=0\ ,
\eqn\xxfour
$$
where $z^1, z^2$ are complex coordinates of 
the p-brane and $s$ is a complex coordinate
transverse to the brane.  The rotation group
$SU(3)$ acts on the coordinates $z^1, z^2, s$.
Choosing the complex structure in a way 
similar to that of the previous case, 
the  spinor singlets of $SU(3)$ satisfy 
the same conditions as in \susyq.  Using 
these two
conditions together with the supersymmetry
 projection of the p-brane, one concludes
that a generic worldvolume soliton will 
preserve $1/8$  of the bulk supersymmetry.

These worldvolume solitons correspond to
 intersecting brane configurations which are
magnetic duals (in the sense of [\gppkt]) to 
the intesecting brane configurations associated
with the solitons of the previous case. A 
typical example is the M-theory configurations
of three M-5-branes pairwise intersecting 
on 3-branes and altogether at a string. Again here
the intersection is at $SU(3)$ angles. However
 for simplicitly we give the orthogonally
intersecting configuration which is 
$$
\eqalign{
{\rm (i)\, M-5:}&\qquad 0,\,1,\,2,\,3,\,4,\,5
\cr 
{\rm (ii)\, M-5:}&\qquad 0,\,1,\,2,\,3,\,-,\,-,\,6,\,7
\cr
{\rm (iii)\, M-5:}&\qquad 0,\,1,\,-,\,-,\,4,\,5,\,6,\,7\ .}
\eqn\intfive
$$
  This
configuration is magnetic dual to the 
configuration of three M-2-branes intersecting
on a 0-brane \inttwo. The degree four 
calibration associated with  a M-5-brane describes a
string worldvolume soliton. From the  
perspective of one of the M-5-branes involved in
the intersection \intfive, say the 
first one,  the string is along the 
directions $0,1$, the
complex coordinates $z^1, z^2$ span 
the directions $2,3,4,5$ and $s$ is along the directions
$6,7$.

\section{SU(4) K\"ahler calibrations}

There are three K\"ahler calibrations 
associated with the group $SU(4)$. These are a
degree two calibrations in $\bC^8$, 
degree four calibrations in $\bC^8$ and degree six
calibrations in $\bC^8$.

The worldvolume soliton of a p-brane
 described by degree two K\"ahler calibration in
$\bC^4$  is 
 the zero locus
of the holomorphic functions
$$
\eqalign{
F^1(s^1,s^2,s^3,z)&=0
\cr
F^2(s^1,s^2,s^3, z)&=0
\cr
F^3(s^1,s^2,s^3, z)&=0\ ,}
\eqn\xxfive
$$
where $z$ is a complex worldvolume coordinate
 and $s^1, s^2, s^3$ are complex coordinates
transverse to the p-brane. Next suppose that 
$z=x^{p-1}+ix^p$, $s^1=y^1+iy^2$,
 $s^2=y^3+iy^4$ and $s^3=y^5+iy^6$ ($y^i=X^{p+i}$).  The
$SU(4)$ rotation group acts on 
$s^1, s^2, s^3, z$ and with this choice of complex structure
the spinor
singlets of
$SU(4)$ satisfy
$$
\eqalign{
\Gamma_{p-1}\Gamma_p\Gamma_{p+1} \Gamma_{p+2}\epsilon&=-\epsilon
\cr
\Gamma_{p-1}\Gamma_p\Gamma_{p+3} \Gamma_{p+4}\epsilon&=-\epsilon
\cr
\Gamma_{p-1}\Gamma_p\Gamma_{p+5} \Gamma_{p+6}\epsilon&=-\epsilon\ .}
\eqn\susyqq
$$
We then choose  the base point
$y_0$ as in the $SU(2)$ case above so 
that the projector of the p-brane is given as
in \susyoneii. Using the p-brane projection
 operator together with \susyqq, we find that the
proportion of the bulk supersymmetry preserved is $1/16$ as expected.
As in the
$SU(2)$ case the above projections can 
be rewritten as the projections of four p-branes
intersecting on a (p-2)-brane.

 An example of such worldvolume soliton
 is the one that is associated with the bulk
configuration  of four M-2-branes  intersecting on a 0-brane, {\sl
i.e.}
$$
\eqalign{
{\rm (i)\, M-2:}&\qquad 0,\,1,\,2,\,-,\,-,\,-,\,-,\,-,\,-
\cr
{\rm (ii)\, M-2:}&\qquad 0,\,-,\,-,\,3,\,4,\,-,\,-,\,-,\,-
\cr
{\rm (iii)\, M-2:}&\qquad 0,\,-,\,-,\,-,\,-,\,5,\,6\,\,-,\,-
\cr 
{\rm (iv)\, M-2:}&\qquad 0,\,-,\,-,\,-,\,-,\,-,\,-\,\,7,\,8 .}
\eqn\inttwoo
$$
 From the perspective of one of the 
four 2-branes, say the first one, $z$ spans the
worldvolume directions $1,2$ and  $s^1, s^2, s^3$ span the
worldvolume directions $3,4,5,6, 7,8$ 
of the other three M-2-branes. The explicit solutions
that
we have given for the case of three
 M-2-branes intersecting on a 0-brane can be easily
generalize to this case and we shall not repeat the analysis here.

Next we shall consider the case of 
degree four K\"ahler calibrations in $\bC^4$. The
worldvolume soliton of a p-brane associated with such 
calibration is described by the
zero locus of two holomorphic functions
$$
\eqalign{
F^1(z^1,z^2, s^1,s^2,)&=0
\cr
F^2(z^1,z^2, s^1,s^2)&=0 \ ,}
\eqn\xxsix
$$
where $z^1,z^2$ are two complex worldvolume
 coordinates and $s^1, s^2$ are two  complex
coordinates transverse to the p-brane.  The 
spinor singlets of $SU(4)$ satisfy the same
conditions \susyqq\ of the previous case 
for a similar choice of complex structure. The
proportion of the supersymmetry preserved 
is $1/16$ of the bulk supersymmetry.

>From the bulk perspective there are 
five intersecting brane configurations that
correspond to this worldvolume solitons. 
 These are two M-5-branes intersecting at
a string, two IIA and IIB NS 5-branes 
intersecting at a string, two D-5-branes
intersecting at a string and two
D-4-branes intersecting at a 0-brane.  All these
configurations are related by 
reduction from M-theory to IIA and T-duality 
from IIA to IIB.  So let us consider the case of two intersecting D-4-branes
at a 0-brane. From the perspective of one of the D-4-branes, the 0-brane
soliton is described by a  degree four
K\"ahler calibration with $z^1,z^2$ 
the worldvolume coordinates of the chosen
brane and $s^1,s^2$ the worldvolume coordinates of the other.
However there is a puzzle, it is 
well known that when the intersection of two D-4-branes is
orthogonal the proportion of the
supersymmery preserved by the 
configuration is $1/4$ of the bulk. 
This is unlike the cases that we have studied
previously where the orthogonally 
intersecting configuration and the intersecting
configuration at $SU(n)$ angles
 preserved the same proportion of supersymmetry.  
One resolution of the  puzzle is that worldvolume soliton
that is associated with the orthogonally 
intersecting configuration does
not utilize the full $SU(4)$ group of
 rotations of the tangent bundle of the
submanifold.  We shall give such solutions in section 8. Such solutions
though are singular at the intersection.

The last case is that of worldvolume
 solitons of p-branes associated with K\"ahler
calibrations of degree six in $\bC^4$.  
This requires that $p\geq 6$ and with at least
two transverse directions. These 
solitons are given by the zero locus of the
holomorphic function
$$
F(s, z^1,z^2,z^3)=0
\eqn\xxseven
$$
where $z^1,z^2,z^3$ are complex worldvolume 
coordinates of the p-brane and $s$ is
a complex transverse coordinate.  The 
investigation of the properties of these
solitons, like supersymmetry, is similar 
to that of the previous cases and we shall
not pursue this further here.

\chapter{Static Solutions and Special Lagrangian Calibrations}

The special Lagrangian calibrations (SLAG)
 are closely related to the
K\"ahler ones which we have investigated
 in the previous sections.
The contact set of a SLAG calibration 
is $G(\phi)=SU(n)/SO(n)$. Therefore a generic
worldvolume soliton associated with a SLAG calibration preserves
$1/2^{n}$ of the supersymmetry of the bulk.

To describe the solutions of Nambu-Goto 
action associated with SLAG calibrations, we again
choose the static gauge
\staticgauge\ of a p-brane as in the 
case of K\"ahler calibrations. The SLAG calibration  is
then a map $y=\{y^i; i=1,\dots, n\}$ from
$\bE^{n}\subset
\bE^{(p,1)}$ with coordinates $\{x^i;i=1,\dots,n\}$ into
$\bE^{n} \subset \bE^{(9,1)}$.  In this
 gauge, the conditions on $y$ required by the SLAG
calibration [\cali, \calib] are
$$
y^i=\partial_i f(x)\ ,
\eqn\calionea
$$
and
$$
{\rm Im} \big({\rm det} (\delta_{ij}+i\partial_i\partial_j f)\big)=0\ , 
\eqn\aaaaaa
$$
where $f$ is a real function of the 
coordinates $\{x^i;i=1,\dots,n\}$ and $\partial_i
f={\partial\over\partial x^i} f$. Some 
examples of SLAG calibrations have
been given in [\cali, \calib] and we shall not repeat them here.

\section{SU(2) SLAG calibrations}

The $SU(2)$ SLAG calibrations are the
 same as the $SU(2)$ K\"ahler ones.
This is because the calibration form in this case is
$$
\phi=dz^1\wedge dz^2+d\bar z^1\wedge d\bar z^2
\eqn\wwone
$$
which is the K\"ahler form of another
 complex structure $J$ on $\bE^4$. Therefore
this SLAG calibration is a K\"ahler 
calibration with respect to $J$. 
So the investigation of the properties of the corresponding
solutions of the Born-Infeld field 
equations is the same as that described in section 5.1   
for the corresponding K\"ahler calibration.

\section{SU(3) SLAG calibrations}

The $SU(3)$ SLAG calibration is a degree 
three calibration in $\bE^6$.
To interprete this calibration as a soliton 
of a p-brane, we take three $\{x^i; i=1,2,3\}$ of
the coordinates of $\bE^6$ to be  worldvolume
 directions of a p-brane ($p>2$) and the
other three $\{y^i; i=1,2,3\}$ to be transverse to it. The
$SU(3)$  rotations act on $\bE^6$ with 
complex coordinates $z^i=x^i+iy^i$.
 The supersymmetry  projections of
$SU(3)$ rotations of the contact set 
can be easily found using the above choice of complex
structure of the SLAG calibration 
(compare with \susyq).  A straightforward computation
reveals that the supersymmetry preserved
 by such a soliton is $1/8$ of the bulk
supersymmetry; $1/4$ of the supersymmetry 
is broken by the $SU(3)$ rotations and another
$1/2$ is broken by the supersymmetry 
projection associated with the p-brane.

 The simple example of a bulk configuration 
which is associated to the solitons of the
$SU(3)$ SLAG calibration is that of three
 M-5-branes  intersecting on a membrane, {\sl
i.e.}
$$
\eqalign{
{\rm (i)\, M-5:}&\qquad 0,\,1,\,2,\,3,\,4,\,5,\,-,\,-,\,-
\cr
{\rm (ii)\, M-5:}&\qquad 0,\,1,\,2,\,-,\,-,\,5,\,6,\,7,\,-
\cr
{\rm (iii)\, M-5:}&\qquad 0,\,1,\,2,\,-,\,4,\,-,\,6\,\,-,\,8
\ .}
\eqn\inttwooo
$$
>From the perspective of one of the three M-5-branes, say the
 first one,  this bulk
configuration is associated to a 2-brane worldvolume 
soliton in the directions $0,1,2$.
The transverse scalars $(y^1, y^2, y^3)$, 
($y^i=X^{p+i}$), of the worldvolume soliton depend on the worldvolume
coordinates $x^3, x^4, x^5$.   
It is straightforward to check that this 
configuration of M-5-branes  preserves $1/8$ of the
bulk supersymmetry. The associated 
supersymmetry projections are identical with
those of the 2-brane worldvolume soliton. 
Reducing this configuration to IIA theory along
$x^1$ and T-dualizing to IIB along $x^2$,  
we find that the same calibration
describes a string soliton on a D-4-brane
 and a 0-brane soliton on a D-3-brane,
respectively.

The above interpretation of the 
worldvolume solitons of the $SU(3)$ SLAG calibrations is not 
unique. This calibration can also be 
interpreted as the soliton of two
M-5-branes intersecting on a 2-brane
 at $SU(3)$ angles.  We remark though that if the two
M-5-branes are  brought in an othogonal 
position  all supersymmetry will break.

\section{SU(4) and SU(5) SLAG calibrations}

The investigation of the worldvolume 
solitons of the $SU(4)$ and $SU(5)$
 SLAG calibrations is similar to the 
one described above for the worldvolume
solitons of the $SU(3)$ SLAG calibrations, 
so we shall not present these cases
in detail. Here we shall
present two examples of bulk configurations
 that can be associated to these worldvolume
solitons the following: (i) A bulk 
configurations associated to $SU(4)$ SLAG calibration 
and preserving $1/16$ of the supersymmetry is
$$
\eqalign{
{\rm (i)\, M-5:}&\qquad 0,\,1,\,2,\,3,\,4,\,5,\,-,\,-,\,-,\,-
\cr
{\rm (ii)\, M-5:}&\qquad 0,\,1,\,-,\,-,\,4,\,5,\,6,\,7,\,-,\,-
\cr
{\rm (iii)\, M-5:}&\qquad 0,\,1,\,-,\,3,\,-,\,5,\,6\,\,-,\,8,\,-
\cr
{\rm (iv)\, M-5:}&\qquad 0,\,1,\,-,\,3,\,4,\,-,\,6\,\,-,\,-,\,9
\ .}
\eqn\intfour
$$
Another configuration can be found by placing 
the third M-5-brane above in the
directions $0,1,2, 5, 7, 8$.  The soliton 
is a string on the M-5-brane and lies in the
directions
$0,1$.
(ii) A bulk configurations associated to $SU(5)$ SLAG calibration and 
preserving $1/32$ of the
supersymmetry  is
$$
\eqalign{
{\rm (i)\, M-5:}&\qquad 0,\,1,\,2,\,3,\,4,\,5,\,-,\,-,\,-,\,-,\,-
\cr
{\rm (ii)\, M-5:}&\qquad 0,\,-,\,-,\,3,\,4,\,5,\,6,\,7,\,-,\,-,\,-
\cr
{\rm (iii)\, M-5:}&\qquad 0,\,1,\,-,\,3,\,-,\,5,\,-\,\,7,\,-,\,9,\,-
\cr
{\rm (iv)\, M-5:}&\qquad 0,\,-,\,2,\,-,\,4,\,5,\,6\,\,-,\,8,\,-,\,-
\cr
{\rm (v)\, M-5:}&\qquad 0,\,1,\,2,\,3,\,-,\,-,\,-\,\,-,\,-,\,9,\,10
\ .}
\eqn\inttthreee
$$
We remark that the M-5-branes can
 be placed at different directions from
those indicated above still 
leading to five M-5-branes intersecting at a
0-brane. The worldvolume soliton is a 0-brane on the M-5-brane.

Apart from the interpretation given above, 
$SU(4)$ and $SU(5)$ SLAG calibrations
 solitons can also be associated with 
two M-5-branes
intersecting at $SU(4)$ and $SU(5)$
 angles on a 1-brane and a 0-brane, respectively.
The latter case may be of interest 
since it is suitable for describing M-5-brane
junctions but this will not be investigated further here. The bulk configuration
of two orthogonally intersecting 
M-5-branes at a string preserves $1/4$ of the supersymmetry
and as we shall show later that 
there is a singular worldvolume 
soliton associated with
it.  However the bulk configuration
 of two orthogonal M-5-branes intersecting at a
0-brane breaks all the supersymmetry 
and therefore for such configuration to exist
the intersection should occur at $SU(5)$ angles.

\chapter{Exceptional calibrations}

\section {$G_2$ calibrations}

The investigation of the solutions of 
the Born-Infeld action associated with the group $G_2$
is similar to that of the K\"ahler and
 SLAG calibrations. However in
this case, there are not seem to be a 
straightforward interpretation of these
calibrations in terms of
intersecting M-branes. It is tempting 
though to suggest that these calibrations
are associated with intersecting 
M-5-branes at $G_2$ angles on a string. Supposing
this, we take the M-5-brane to lie
 in the directions $0,1,2,6,7,8$ and the string to lie
in the directions $0,8$. For the 
degree three calibration, we take $\bE^7$ to span the
directions $1,2,3,4,5,6,7$.  The spinor singlets under $G_2$ satisfy
$$
G_{mn}\epsilon=0
\eqn\singleone
$$
where
$$
G_{mn}=\Gamma_{mn}+{1\over4} \star \varphi_{mn}{}^{pq} \Gamma_{pq}
\eqn\wwtwo
$$
are the generators of $G_2$, 
$m,n,p,q=1,\dots,7$, and $\star \varphi$ is the
Hodge dual of the structure constants 
$\varphi$ of the octonions which we have chosen as
$$
\varphi_{123}=\varphi_{246}=\varphi_{435}=\varphi_{516}
=\varphi_{572}=\varphi_{471}=\varphi_{673}=1
\eqn\wwthree
$$ 
The condition \singleone\ on $\epsilon$ yields the supersymmetry
projections
$$
\eqalign{
\Gamma_{1346}\epsilon&=\epsilon
\cr
\Gamma_{2356}\epsilon&=\epsilon
\cr
\Gamma_{4567}\epsilon&=\epsilon\ .}
\eqn\gtwoproj
$$
There are many equivalent ways to present 
these projections for the above choice of
$\varphi$.
The consistency of these projections 
together with the projector associated
with the M-5-brane reveals that a generic
 $G_2$ calibration preserves $1/16$
of the bulk supersymmetry.
For the degree four calibration, we again 
take $\bE^7$ to span the directions
$1,2,3,4,5,6,7$. The supersymmetry projectors
 associated with $G_2$ are as in \gtwoproj. The
supersymmetry preserved is also
$1/16$ of the bulk.

\section {${\rm Spin}(7)$ calibrations}

The ${\rm Spin}(7)$ calibrations are 
degree four calibrations in $\bE^8$ and a
bulk interpretation is as two M-5-branes 
at ${\rm Spin}(7)$ angles intersecting on a string.
As in the $G_2$ case above, we take one of 
the M-5-branes to lie in the directions
$0,1,2,6,7,9$, the string to lie in the
 directions $0,9$ and the calibration to take place
in the directions $1,2,3,4,5,6,7,8$. 
 The spinor singlets under ${\rm Spin}(7)$ satisfy
$$
G_{IJ}\epsilon=0
\eqn\singletwo
$$
where
$$
G_{IJ}={3\over4} \big(\Gamma_{IJ}+{1\over6}  \Omega_{IJ}{}^{KL} \Gamma_{KL}\big)
\eqn\wwfive
$$
are the generators of ${\rm Spin}(7)$, $I,J,K,L=1,\dots,8$, and 
$$
\eqalign{
\Omega_{mnp8}&=\varphi_{mnp}
\cr
\Omega_{mnpq}&=\star\varphi_{mnpq}}
\eqn\sone
$$
is the ${\rm Spin}(7)$-invariant self-dual 4-form.
The supersymmetry projections associated
with ${\rm Spin}(7)$ acting on $\bE^8$ 
spanned by the above directions are
$$
\eqalign{
\Gamma_{1346}\epsilon&=\epsilon
\cr
\Gamma_{2356}\epsilon&=\epsilon
\cr
\Gamma_{4567}\epsilon&=\epsilon
\cr
\Gamma_{1238}\epsilon&=\epsilon\ .}
\eqn\spinproj
$$

The first three projections are similar to 
those of $G_2$. Using the last two projections,
we find that
$$
\Gamma_{12345678}\epsilon=\epsilon\ .
\eqn\stwo
$$
This implies that if $\epsilon$ is anti-chiral 
in the 8-dimensional sense all supersymmetry
is broken\foot{We remark that this depends 
on the orientation of $\bE^8$. If the other
orientation is chosen, then chiral
 representation does not have singlets under
${\rm Spin}(7)$.} The consistency of 
these projections together with the projector associated
with the M-5-brane  reveals that 
a generic ${\rm Spin}(7)$ calibration preserves $1/32$
of the bulk supersymmetry.

We summarize some of our results in the sections five-seven in the table below. 
 \vskip 0.5cm
$$
\vbox{\settabs 3\columns 
\+ {\rm Calibration}& {\rm Contact}  & {\rm Supersymmetry}\cr
\+{\rm K\"ahler}&${SU(n)\over S\big(U(p)\times U(n-p)\big)}$& $~~~2^{-n}$\cr
\+ {\rm SLAG}&${SU(n)\over SO(n)}$& $~~~2^{-n}$\cr
\+ $G_2$&${G_2\over SO(4)}$& $~~~2^{-4}$\cr
\+ {\rm Spin(7)}&${{\rm Spin}(7)\over H}$& $~~~2^{-5}$\cr}
$$
\vskip 0.2cm
\noindent{\bf Table 2: Calibrations and Supersymmetry}\ \
This table contains
the type of calibration, the associated 
contact set and the proportion of the supersymmetry
preserved by the calibration.
\noindent

\chapter{Singular Solutions and Piecewise Planar Branes}

The Dirac-Born-Infeld field equations
 admit a large class of
solutions for which the transverse 
scalars $y$ are piece-wise linear functions
 and the Born-Infeld field $F$ is 
piece-wise constant. In this class of solutions, we
can also include double valued solutions
 like for example the configurations for which
their graph consisits of two or more
 geometrically intersecting planes in $\bE^{(9,1)}$.
 Some of these solutions are limits of the solitons
 discussed in the previous sections. The
bulk interpretation of such solutions is that of
 planar branes intersecting at angles with  a
non-vanishing  Born-Infeld field. The 
singularities of these solutions are at the
intersections where the transverse scalars and their
first derivatives  are discontinuous. We shall call such solutions
of the Dirac-Born-Infeld field equations \lq singular solutions'.

The investigation of the supersymmetry 
preserved by a singular solution with
vanishing Born-Infeld field is very 
similar to that of two or more planar branes
placed in the ten-dimensional Minkowski
 background. Such analysis has already
been done in [\kallosh] and we shall
 not repeat it here.  Instead, we shall give
one example to illustrate the way that 
such a computation can be done.
For this we shall consider singular 
solutions of the M-2-brane action.
  Let $x^1,
x^2$ be the two spatial wordvolume 
coordinates of a M-2-branes and
$y^1=X^3, y^2=X^4$ two of its transverse scalars. 
 A singular worldvolume solution of the
M-2-brane is
$$
y^i=\cases{c^i\ , \qquad {\rm for}\quad x^1\cdot x^2>0\cr
x^i\ , \qquad {\rm for}\quad x^1\cdot x^2<0\ ,}
\eqn\sinone
$$
where $\{c^i; i=1,2\}$ are constants.
  The
supersymmetry conditions associated with this solution are
$$
\eqalign{
\Gamma_0 \Gamma_1 \Gamma_2\epsilon&=\epsilon\ , \qquad x^1\cdot  x^2>0
\cr
{1\over2}\Gamma_0 (\Gamma_1+\Gamma_3) (\Gamma_2+\Gamma_4)\epsilon&=\epsilon
\ , \qquad x^1\cdot x^2<0\ .}
\eqn\sintwo
$$
Using the first projector the second one can be rewritten as
$$
\Gamma_0 \Gamma_3 \Gamma_4\epsilon=\epsilon\ .
\eqn\sthree
$$
Therefore the supersymmetry preserve is $1/4$ of 
the bulk. The projectors
of this solution are those of two M-2-branes 
intersecting at a 0-brane.

We can easily include piece-wise constant 
Born-Infeld fields to the above singular solutions.
If the Born-Infeld is everywhere constant, 
then it is straightforward to see that its
inclusion does not break any additional 
supersymmetry.  However if it is piece-wise constant,
additional supersymmetry may be broken (see [\kallosh]).

\chapter{Solutions with only Born-Infeld fields}

A consistent truncation of the Dirac-Born-Infeld action
 is to fix the location of the p-brane,
as in section 3.2,
and reduce it to a Born-Infeld action.
Choosing a Lorentz frame in the bulk adopted to 
the p-brane, the supersymmetry condition,
given also in section 3.2, becomes
$$
e^{-{ a\over 2}} \Gamma 
   e^{ a\over 2}\epsilon=\epsilon
\eqn\bipra
$$
since $S=1$; the projector 
$\Gamma$ is a product of (constant) Gamma matrices.
As it has already been mentioned in section 3.2 
 linearizing this condition
in the Born-Infeld fields, it becomes the
 familiar supersymmetry condition of the
maximally supersymmetric Maxwell multiplet
 in (p+1) dimensions together with
the projection associated with the planar p-brane.
So at least in the linear approximation, 
Born-Infeld fields that satisfy the self-duality
condition and its generalizations, 
like for example the Einstein-Yang-Mills ($SU(n)$)
condition or the
$Sp(k)$ condition, preserve 
a proportion of the spacetime
supersymmetry. We remark that the
worldvolume solitons involving only
 Born-Infeld fields have the bulk interpretation of branes
within branes.

A general analysis for the conditions 
that $Y$ (see \tana) should satisfy for \bipra\ to
admit not trivial solutions has been 
already given in section 3.2. Here we shall
examine some special cases. First we shall 
consider the case where $Y$ is self-dual 2-form
in the directions 1,2,3,4 of the p-brane ($p\geq 4$). 
If this is the case, then the singlets
of the rotation $e^a$ satisfy
$$
\Gamma_{1234}\epsilon=-\epsilon\ .
\eqn\selfproj
$$
We remark that the same condition on $\epsilon$ 
is imposed by a self-dual Born-Infeld field
in the linearized limit. 
If $\epsilon$ is such a singlet, then the 
supersymmetry condition reduces to
$$
\Gamma \epsilon=\epsilon\ .
\eqn\pbraneproj
$$
The compactability conditions of the two
 projections \selfproj\ and \pbraneproj\
determine the proportion of the bulk 
supersymmetry preserved by the configuration.
It turns out that self-dual configurations
 on all D-p-branes $p\geq 4$ preserve
$1/4$ of the bulk supersymmetry. 
It remains
to express the self-duality condition on 
$Y$ in terms of a condition on $F$.
It turns out that if $Y$ is self-dual so 
is $F$ as it can be easily seen using the
 the identity
$$
Y_{ac} Y_{db} \delta^{cd}= -{1\over4} Y_{cd} Y^{cd} \delta_{ab}
\eqn\sten
$$
and the
relation $F={\rm tan} Y$, where $a,b,c,d=1,\dots, 4$. 
 Moreover a short calculation reveals 
that $F$ satisfies the field equations of the Born-Infeld action
as a consequence of the self-duality condition and the
 Bianchi identity. 

A simple explicit example with a self-dual 
Maxwell field is obtained by setting
$$
F_{ab}=\partial_a A_b-\partial_b A_a
\eqn\solaaaa
$$
and 
$$
A_a= I_a{}^b \partial_b H
\eqn\solbbbb
$$
where $I$ is a constant complex 
structure of $\bE^4$, $I_a{}^b I_b{}^c=-\delta_a{}^c$,
with anti-self-dual K\"ahler form and 
$H$ a harmonic function on $\bE^4$. This
worldvolume solution has several bulk 
interpretations depending on the context that
it is used. One example is that of a D-0-brane within a D-4-brane.

Other conditions on $Y$ which lead to 
preservation of some of the supersymmetry
of the bulk have already been  mentioned 
is section 3.2. However there does not seem to be
such a simple relation between these
 conditions of $Y$ and those on $F$. For example,
if
$Y$ satisfies the K\"ahler-Yang-Mills 
condition (i.e. $Y$ is in the Lie algebra $SU(k)$), one
can easily see that
$F$ does not satisfy the same condition. We remark,
 however, that if $Y$ is tri-hermitian
(i.e.
$Y$ is in the Lie algebra 
$Sp(k)$), then
$F$ is tri-hermitian as well.  Such condition 
has been studied [\garypaul] where it
was found  that the supersymmetry preserved 
$3/16$ of the bulk. Explicit solutions can be
easily constructed by  superposing those 
of \solbbbb\ using the  method developed in [\gptf].

\chapter{Solitons with Born-Infeld fields}

Solitons of the Dirac-Born-Infeld field 
equations of a p-brane that involve both non-vanishing
scalars and Born-Infeld fields are those 
that arise at the boundary of one brane ending on
another.  Such solitons preserve $1/4$ of 
the bulk supersymmetry or less. We have not been
able to find a calibration-like construction
 to deal with these solitons. So we shall use the
bulk picture to determine the existence of
 such solutions and we shall give some new
examples. It is convenient to categorize 
them according to the supersymmetry that they
preserve.

\section{Solitons preserving $1/4$ of the supersymmetry}

Most of these solitons have already being 
found. These include, (i) the 0-brane
soliton on all D-branes due to a fundamental 
string ending on
them [\gary], (ii) the 0-brane soliton on the 
IIB NS-5-brane due to a D-string ending on it,
(iii)the self-dual string soliton [\howeb] on 
the M-5-brane due to a membrane ending on
it, and (iv) the 2-brane soliton [\howeb] of 
the IIB NS-5-brane due to a three brane
ending on it.  These solitons are related 
using T-and S-duality transformations of
the associated bulk theories [\gpapas]. 
Apart from these soliton solutions which 
are charged with respect to the standard
2-form Born-Infeld field, there are other 
solitons which are charged with respect to other
p-form fields. One example is the domain 
wall solutions on the IIB D-5-and NS-5-branes
[\bergpap]. We remark that in all the 
above solitons, the superymmetry projectors
are those of the associated bulk 
configuration. For example the supersymmetry
projectors of the self-dual string 
soliton on the M-5-brane are those on
the M-5-brane and the M-2-brane.The  
energy bounds for some of the above solutions have been
investigated in [\gomis] and the geometry 
of their moduli spaces have been studied in
[\gpapasb].

\section{Solitons preserving $1/8$ of the supersymmetry}

There are many such solitons. This can be seen either by studying
the charges that appear in the various worldvolume supersymmetry algebras or by
examinning the bulk intersecting
 brane configurations.
We shall not present a complete account 
of all such configurations.
Instead we shall investigate a class of 
electrically charged solutions
associated with D-branes. Let $F$ be the
 Born-Infeld field of a D-p-brane and $\{y^i;
i=1,\cdots, 9-p\}$ be the transverse scalars.
 For electically charged solutions, we use the
ansatz
$$
\eqalign{
F_{0a}&=-\partial_a\phi\ ,\quad a=1,\dots,k
\cr
y^1&=\phi\ ,
\cr
y^m&=y^m(x)\quad m=2,\dots,9-p }
\eqn\aa
$$
A standard computation using the results of 
[\gary] reveals that the field equations of the
D-p-brane become
$$
\eqalign{
g^{ab} \nabla_a \partial_b\phi&=0
\cr
g^{ab} \nabla_a \partial_b y^m&=0}
\eqn\deq
$$
where
$$
g_{ab}=\delta_{ab}+ \partial_a y^m \partial_b y^n \delta_{mn}
\eqn\ind
$$
is the induced metric and $\nabla$ is the 
Levi-Civita connection of $g$.
The second equation in \deq\ can be 
solved by taking the maps $\{y^m\}$ to describe a
degree $k$ calibration in $\bE^{8+k-p}$. 
The first equation in \deq\ is just the harmonic
function condition on the calibrated surface.

The equation \deq\ can be easily 
solved if the associated calibration is K\"ahler.
In this case the induced metric $g$ on
 the calibrated manifold is K\"ahler. So the
first equation of \deq\ can be solved by 
taking $\phi$ to be the real part of a holomorphic
function $h$, {\sl i.e.}
$$
\phi=h(z)+\bar h(z)\ .
\eqn\tone
$$
There are many bulk configuration that  
correspond to the above solutions by allowing
for different calibrations.
One example is the 0-brane worldvolume 
soliton on the D-3-brane corresponding
to the intersection
$$
\eqalign{
{\rm (i)\, D-3:}&\qquad 0,\,1,\,2,\,3,\,-,\,-,\,-
\cr
{\rm (ii)\, D-3:}&\qquad 0,\,1,\,-,\,-,\,4,\,5,\,-
\cr
{\rm (iii)\, F-1:}&\qquad 0,\,-,\,-,\,-,\,-,\,-,\,6\ .}
\eqn\ttwo
$$
The associated calibration is a degree two 
$SU(2)$ K\"ahler calibration in the directions
$2,3,4,5$ and $y^1=X^6$. The proportion of 
the supersymmetry preserved is $1/8$.
We shall present a complete discussion of 
the worldvolume solitons that preserve
$1/8$ of the supersymmetry or less elsewhere.

\chapter{Conclusions}

We have investigated  the Killing spinor 
equations associated with the kappa-symmetry
transformations of the worldvolume brane 
actions and shown that all the worldvolume
solitons associated with calibrations are 
supersymmetric. The proof has been based
on the properties of the contact set of a calibration. Then we
have presented a bulk interpretation of all 
the calibrations in terms of intersecting branes.
Next we have examined other supersymmetric 
worldvolume solutions, like
singular solutions and solutions 
involving only Born-Infeld fields. Finally, we found
new worldvolume solutions of the 
Dirac-Born-Infeld action. In particular, we present a 0-brane
worldvolume solution with the bulk 
interpretation of two  intersecting D-3-brane with a
fundamental string ending on them. 
Such solution preserves $1/8$ of supersymmetry.

So far we have considered solutions of the Nambu-Goto-type of actions
which are embeddings into the ten- or eleven-dimensional Minkowski
spacetimes. Alternatively, we can take as spacetime any solution
of  ten- or eleven-dimensional supergravities. In particular, we can take
${\cal M}_{(10)}=M_k\times \bE^{(9-k, 1)}$ 
or ${\cal M}_{(11)}=M_k\times \bE^{(10-k, 1)}$
where $M_k$ is a manifold with the special holonomies of table 1. Since for all these
manifolds the Ricci tensor vanishes, the above 
spaces are solutions of
the supergravity field equations by setting
 the rest of the fields either
zero or constant. One can then consider
 solutions of the Nambu-Goto-type
of actions for the spacetimes given as above (see also [\beckera, \beckerb]). 
There is a natural definition
of a calibration in the manifolds with special 
holonomy using the covariantly constant
forms (see table 1). It turns out that such solutions
of the Nambu-Goto action associated with these calibrations
 preserve the same proportion of
supersymmetry as those we have studied for Minkowski spacetimes.

We have shown that calibrations with 
contact set $G/H$ are associated with
intersecting branes at $G$-angles. It 
is well known that if $G=Sp(2)$ there are
supergravity solutions which have the
 interpretation of two M-5-branes
intersecting at a string at $Sp(2)$ angles. 
There does not seem to be though
a calibation with contact set $Sp(2)/H$, 
or more general $Sp(n)/H$, based
on a calibrating form. However, it is 
straightforward to define one in terms
of spinors. The relevant computation has 
already been done in [\garypaul]. It would be of
interest to find the corresponding 
calibration based on forms. One option of such calibration
is that for which the tangent spaces 
are quaternionic planes in $\bE^{4n}$.

It would be of interest to investigate
 further the Dirac-Born-Infeld type of actions. In
particular, there may be a generalization 
of calibrations which also involves  non-vanishing
Born-Infeld field. Unlike for calibrations, 
in this case it is not clear which geometric
quantity  is minimized. However, the definition
 of such a generalized calibration in terms of
spinors is straightforward and it has been done 
in section 3. Since it is expected that there
should be a consistent generalization of 
Dirac-Born-Infeld actions involving non-abelian
Born-Infeld field (see [\tseytlin]), there may also exist a
\lq non-abelian' generalization of calibrations.

\vskip 1cm
\noindent{\bf Acknowledgments:}  We thank Jerome Gauntlett for helpful
discussions  and for telling us about related work which will appear
in a forthcoming paper of his with Neil Lambert and Peter West.  G.P.
is supported by a University Research Fellowship from the Royal
Society.

\refout

\bye